\pdfoutput=1

\documentclass[11pt,prd,aps,amssymb,amsmath,tightenlines,showpacs]{revtex4}
\usepackage{graphicx}

\newcommand{\cF}{\ensuremath{\mathcal{F}}}
\newcommand{\half}{\mbox{$\textstyle{\frac{1}{2}}$}}

\begin{document}

\title{Nonlinear eigenvalue problems}

\author{Carl M. Bender$^{a,b}$}\email{cmb@wustl.edu}
\author{Andreas Fring$^b$}\email{a.fring@city.ac.uk}
\author{Javad Komijani$^a$}\email{jkomijani@physics.wustl.edu}
\affiliation{$^a$Department of Physics, Washington University, St. Louis, MO
63130, USA \\
$^b$Department of Mathematical Science, City University London,\\
$\,\,$ Northampton Square, London EC1V 0HB, UK}

\date{\today}

\begin{abstract}
This paper presents a detailed asymptotic study of the nonlinear differential
equation $y'(x)=\cos[\pi xy(x)]$ subject to the initial condition $y(0)=a$.
Although the differential equation is nonlinear, the solutions to this
initial-value problem bear a striking resemblance to solutions to the
time-independent Schr\"odinger eigenvalue problem. As $x$ increases from $x=0$,
$y(x)$ oscillates and thus resembles a quantum wave function in a classically
allowed region. At a critical value $x=x_{\rm crit}$, where $x_{\rm crit}$
depends on $a$, the solution $y(x)$ undergoes a transition; the oscillations
abruptly cease and $y(x)$ decays to $0$ monotonically as $x\to\infty$. This
transition resembles the transition in a wave function that occurs at a turning
point as one enters the classically forbidden region. Furthermore, the initial
condition $a$ falls into discrete classes; in the $n$th class of initial
conditions $a_{n-1}<a<a_n$ ($n=1,\,2,\,3,\,\ldots$), $y(x)$ exhibits exactly $n$
maxima in the oscillatory region. The boundaries $a_n$ of these classes are the
analogs of quantum-mechanical eigenvalues. An asymptotic calculation of $a_n$
for large $n$ is analogous to a high-energy semiclassical (WKB) calculation of
eigenvalues in quantum mechanics. The principal result of this paper is that as
$n\to\infty$, $a_n\sim A\sqrt{n}$, where $A=2^{5/6}$. Numerical analysis reveals
that the first Painlev\'e transcendent has an eigenvalue structure that is quite
similar to that of the equation $y'(x)=\cos[\pi xy(x)]$ and that the $n$th
eigenvalue grows with $n$ like a constant times $n^{3/5}$ as $n\to\infty$.
Finally, it is noted that the constant $A$ is numerically very close to the
lower bound on the power-series constant $P$ in the theory of complex variables,
which is associated with the asymptotic behavior of zeros of partial sums of
Taylor series.
\end{abstract}


\maketitle

\section{Introduction}
\label{s1}
This paper presents a detailed asymptotic analysis of the nonlinear
initial-value problem
\begin{equation}
y'(x)=\cos[\pi xy(x)],\quad y(0)=a.
\label{e1}
\end{equation}
This remarkable and deceptively simple looking differential equation was given
as an exercise in the text by Bender and Orszag \cite{R1}. Since then, it and
closely related differential equations have arisen in a number of physical
contexts involving the complex extension of quantum-mechanical probability
\cite{R2,R3} and the structure of gravitational inspirals \cite{R4}. We will see
that the properties of solutions to this equation are strongly analogous to
those of the time-independent Schr\"odinger eigenvalue problem.

Recall that the Schr\"odinger eigenvalue problem has the general form
\begin{equation}
-\psi''(x)+V(x)\psi(x)=E\psi(x),\qquad\psi(\pm\infty)=0,
\label{e2}
\end{equation}
where $E$ is the eigenvalue. For simplicity, we assume that the potential $V(x)$
has one local minimum and rises monotonically to $\infty$ as $x\to\pm\infty$.
In general this eigenvalue problem is not analytically solvable except for
special potentials [such as the harmonic oscillator potential $V(x)=x^2$].
However, it is possible to find the large-$n$ asymptotic behavior of the $n$th
eigenvalue $E_n$ by using semiclassical (WKB) analysis. To leading order the
large-$n$ asymptotic behavior of the eigenvalues of the two-turning-point
problem may be obtained from the Bohr-Sommerfeld condition
\begin{equation}
\int_{x_1}^{x_2}dx\sqrt{E_n-V(x)}\sim(n+1/2)\pi\quad(n\to\infty),
\label{e3}
\end{equation}
where the turning points $x_1$ and $x_2$ are real roots of the equation $V(x)=
E_n$. This WKB condition determines the eigenvalues implicitly for large $n$. As
an example, for the anharmonic potential $V(x)=x^4$ the large-$n$ asymptotic
behavior of the eigenvalues is \cite{R5}
\begin{equation}
E_n\sim Bn^{4/3}\quad(n\to\infty),
\label{e4}
\end{equation}
where the constant $B$ is given by $B=3\Gamma(3/4)\sqrt{\pi}/\Gamma(1/4)$.

The quantum eigenfunctions $\psi(x)$ exhibit several well known characteristic
features. In the classically allowed region between the turning points ($x_1<x<
x_2$), the eigenfunctions are oscillatory and the eigenfunction corresponding to
$E_n$ has $n$ nodes. In the classically-forbidden regions $x>x_2$ and $x<x_1$
the eigenfunctions decay exponentially and monotonically to zero as $|x|\to
\infty$. Thus, at the turning points the behavior of the eigenfunctions changes
abruptly from rapid oscillation to smooth exponential decay.

\begin{figure}[h!]
\begin{center}
\includegraphics[scale=1.35]{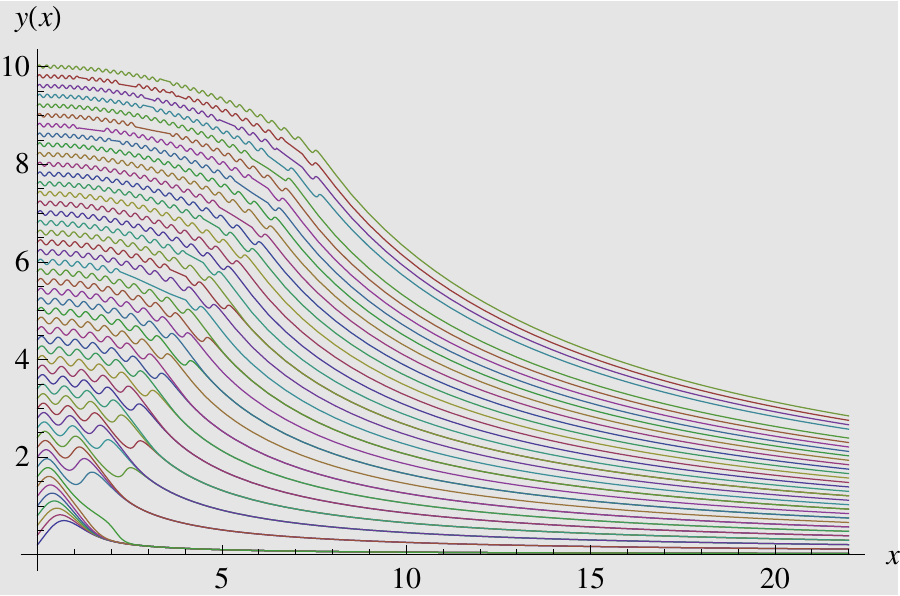}
\end{center}
\caption{Numerical solutions $y(x)$ to (\ref{e1}) for $0\leq x\leq24$ with
initial conditions $y(0)=0.2k$ for $k=1,\,2,\,3,\,\ldots,\,50$. The solutions
initially oscillate but abruptly become smoothly and monotonically decaying. In
the decaying regime the solutions merge into discrete quantized bundles.}
\label{F1}
\end{figure}

The solutions $y(x)$ to the nonlinear differential equation (\ref{e1}) have many
features in common with the solutions $\psi(x)$ to the Schr\"odinger equation
(\ref{e2}). For any choice of $y(0)=a$ the initial slope $y'(0)$ is $1$. As $x$
increases from $0$, $y(x)$ oscillates as shown in Fig.~\ref{F1}. This regime of
oscillation is analogous to a classically allowed region in quantum mechanics.
Note that the number of maxima of the function $y(x)$ in the oscillatory region
increases as $y(0)$ increases. With increasing $x$ the oscillations abruptly
cease, and as $x\to\infty$ the function $y(x)$ then decays smoothly and
monotonically to 0. This kind of behavior resembles that of $\psi(x)$ in a
classically forbidden region.

Figure \ref{F1} reveals that in the decaying regime the curves merge into
quantized bundles. This large-$x$ asymptotic behavior of $y(x)$ can be explained
by using elementary asymptotic analysis. If we seek an asymptotic behavior of
the form $y(x)\sim c/x$ ($x\to \infty$) and substitute this {\it ansatz} into
(\ref{e1}), we find that $c=m+1/2$ ($m=0,\,1,\,2,\,3,\,\ldots$). This is just
the {\it leading} term in the asymptotic expansion of $y(x)$ for large $x$. The
full series has the form
\begin{equation}
y(x)\sim\frac{m+1/2}{x}+\sum_{k=1}^\infty\frac{c_k}{x^{2k+1}}\quad(x\to\infty).
\label{e5}
\end{equation}
The first few coefficients are
\begin{eqnarray}
c_1&=&\frac{(-1)^m}{\pi}(m+1/2),\nonumber\\
c_2&=&\frac{3}{\pi^2}(m+1/2),\nonumber\\
c_3&=&(-1)^m\left[\frac{(m+1/2)^3}{6\pi}+\frac{15(m+1/2)}{\pi^3}
\right],\nonumber\\
c_4&=&\frac{8(m+1/2)^3}{3\pi^2}+\frac{105(m+1/2)}{\pi^4},
\nonumber\\
c_5&=&(-1)^m\left[\frac{3(m+1/2)^5}{40\pi}+\frac{36(m+1/2)^3}{\pi^3}+\frac{945
(m+1/2)}{\pi^5}\right],\nonumber\\
c_6&=&\frac{38(m+1/2)^5}{15\pi^2}+\frac{498(m+1/2)^3}{\pi^4}+\frac{10395(m+1/2)}
{\pi^6}.
\label{e6}
\end{eqnarray}

\subsection{Hyperasymptotic analysis}
\label{ss1A}

A close look at Fig.~\ref{F1} shows a surprising result: Half of the predicted
large-$x$ asymptotic behaviors in (\ref{e5}) appear to be missing. The bundles
of curves shown in Fig.~\ref{F1} correspond only to {\it even} values of $m$. To
explain what has happened to the odd-$m$ bundles, we perform a hyperasymptotic
analysis (asymptotics beyond all orders) \cite{R6}. Let $y_1(x)$ and $y_2(x)$
represent two different curves in the $m$th bundle. Even though they are
different curves they have exactly the same asymptotic approximation as given in
(\ref{e5}). Then $Y(x)\equiv y_1(x)-y_2(x)$ satisfies the differential equation
\begin{eqnarray}
Y'(x)&=&\cos[\pi xy_1(x)]-\cos[\pi xy_2(x)]\nonumber\\
&=&-2\sin\left[\half\pi xy_1(x)+\half\pi xy_2(x)\right]\sin\left[\half\pi xy_1
(x)-\half\pi x y_2(x)\right]\nonumber\\
&\sim&-2\sin\left[\pi\left(m+\half\right)\right]\sin\left[\half\pi
xY(x)\right]\quad(x\to\infty)\nonumber\\
&\sim&-(-1)^m\pi xY(x)\quad(x\to\infty).
\label{e7}
\end{eqnarray}
Thus, we conclude that
\begin{equation}
Y(x)\sim K\exp\left[-(-1)^m\pi x^2\right]\quad(x\to\infty),
\label{e8}
\end{equation}
where $K$ is an arbitrary constant. This calculation shows that while two
different curves in the same bundle have the same asymptotic expansion for large
$x$, they differ by an exponentially small amount. This result explains why no
arbitrary constant appears in the asymptotic expansion (\ref{e5}); the arbitrary
constant appears in the beyond-all-orders hyperasymptotic (exponentially small)
correction to this asymptotic series.

\begin{figure}[h!]
\begin{center}
\includegraphics[trim=32mm 85mm 32mm 85mm,clip=true,scale=1.00]{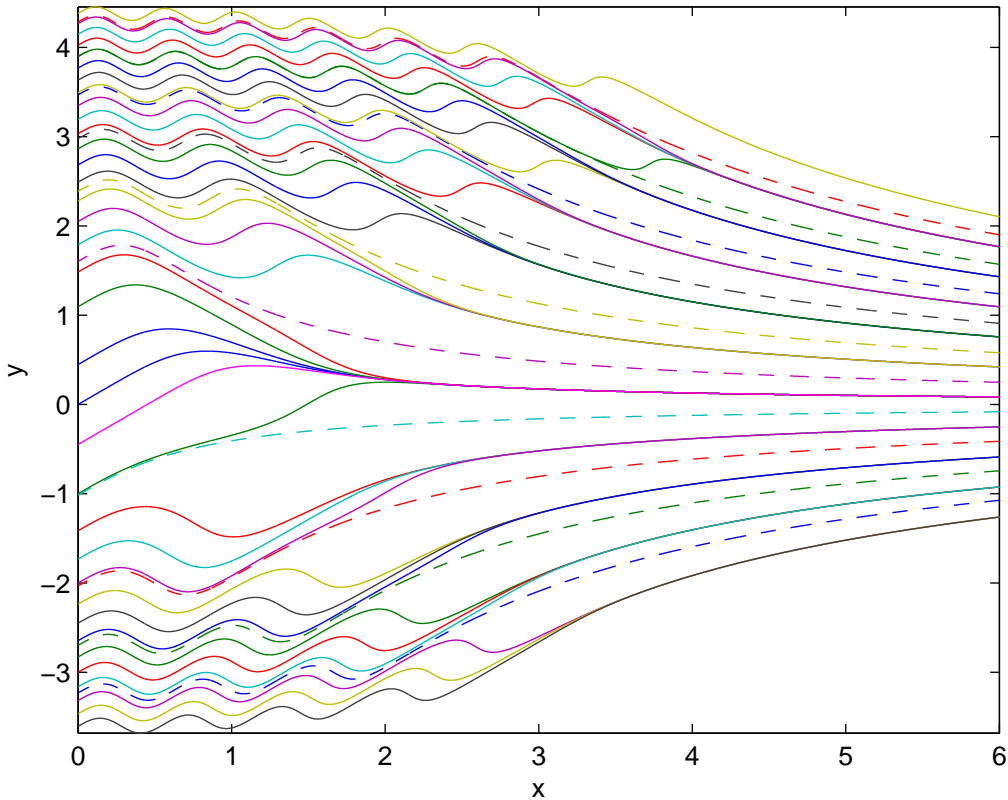}
\end{center}
\caption{Numerical solutions to (\ref{e1}) showing ten separatrix curves, which
cross the $y$ axis at $a_{-3}=-3.231360$, $a_{-2}=-2.698369$, $a_{-1}=-
2.032651$, $a_0=-1.016702$, $a_1=1.602573$, $a_2=2.388358$, $a_3=2.976682$,
$a_4=3.467542$, $a_5=3.897484$, and $a_6=4.284674$.}
\label{F2}
\end{figure}

More importantly, this argument demonstrates that two curves can only be in the
same bundle if $m$ is {\it even}. If $m$ is odd, the two curves {\it move away
from one another} as $x$ increases. Thus, while there is a bundle of infinitely
many curves when $m$ is even, we see that there is a unique and {\it discrete}
curve, called a {\it separatrix}, for the case of odd $m$. The $n$th separatrix,
whose large-$x$ asymptotic behavior is $(2n-1/2)/x$ ($n=1,\,2,\,3,\,\ldots$), is
{\it unstable} for increasing $x$; that is, as $x$ increases, nearby curves
$y(x)$ veer away from it and become part of the bundles above or below the
separatrix. This explains why there are no curves shown in Fig.~\ref{F1} when
$m$ is odd. Ten separatrix curves are shown in Fig.~\ref{F2}.

While the separatrix curves are unstable for increasing $x$, they are stable
for decreasing $x$ and thus it is numerically easy to trace these curves
backward from large values of $x$ down to $x=0$. We treat the discrete point
$a_n$ ($n=1,\,2,\,3,\,\ldots$) at which the $n$th separatrix crosses the $y$
axis as an eigenvalue. The curves $y(x)$, whose initial values $y(0)=a$ lie in
the range $a_{n-1}<y(0)<a_n$, have $n$ maxima. Our objective in this paper is to
determine analytically the large-$n$ asymptotic behavior of the eigenvalues; we
will establish that 
\begin{equation}
a_n\sim A\sqrt{n}\quad(n\to\infty),
\label{e9}
\end{equation}
where $A=2^{5/6}$. The constant $A$ is a nonlinear analog of the WKB constant
$B$ in (\ref{e4}).

Hyperasymptotics also plays a crucial role in quantum theory. Because the
Schr\"odinger differential equation for the eigenvalue problem (\ref{e2}) is
second order, the asymptotic behavior of $\psi(x)$ as $x\to\infty$ contains two
arbitrary constants. However, there is only {\it one} constant $C$ in the WKB
asymptotic approximation
\begin{equation}
\psi(x)\sim C[V(x)-E]^{-1/4}\exp\left[\int^x ds\sqrt{V(s)-E}\right]\quad
(x\to\infty).
\label{e10}
\end{equation}
There is a second constant $D$, of course, but this constant multiplies the
subdominant (exponentially decaying) solution, and thus this constant does not
appear to any order in the WKB expansion. The constant $D$ remains invisible
except at an eigenvalue because only at an eigenvalue does the coefficient $C$
of the exponentially growing solution (\ref{e10}) vanish {\it to all orders} in
the large-$x$ asymptotic expansion, leaving the physically acceptable
exponentially decaying solution
\begin{equation}
\psi(x)\sim D[V(x)-E]^{-1/4}\exp\left[-\int^x ds\sqrt{V(s)-E}\right]\quad
(x\to\infty).
\label{e11}
\end{equation}

\subsection{Organization of this paper}
\label{ss1B}
The principal thrust of the analysis in this paper is an asymptotic study of the
separatrices, which for large $x$ are approximated by the formula in (\ref{e5})
with $m$ odd. Thus, we let $m=2n-1$ and we scale both the independent and
dependent variables in (\ref{e1}):
\begin{equation}
x=\sqrt{2n-1/2}\,t,\quad y(x)=\sqrt{2n-1/2}\,z(t),
\label{e12}
\end{equation}
and let
\begin{equation}
\lambda=(2n-1/2)\pi.
\label{e13}
\end{equation}
The resulting equation for $z(t)$ is
\begin{equation}
z'(t)=\cos[\lambda tz(t)].
\label{e14}
\end{equation}
With these changes of variable, the $n$th separatrix [which behaves like $(2n-
1/2)/x$ as $x\to\infty$] now behaves like $1/t$ as $t\to\infty$. Also, for large
$\lambda$ the turning point (the point at which the oscillations cease and
monotone decreasing behavior begins) is located at $t=1$.

In Sec.~\ref{s2} we begin by examining the differential equation (\ref{e1})
numerically. We then show numerically that for large $\lambda$ the solution
$z(t)$ to the scaled equation (\ref{e14}) that satisfies the initial condition
$z(0)=2^{1/3}$ is oscillatory until $t=1$, at which point it decays smoothly
like $z(t)\sim 1/t$ as $t\to\infty$. We also show that the amplitude of the
oscillations is of order $1/\lambda$ for large $\lambda$. Hence, in the limit
$\lambda\to\infty$ the function $z(t)$ converges to a smooth and nonoscillatory
function $Z(t)$ that passes through $2^{1/3}$ at $t=0$ and through $1$ at $t=1$.
Thus, the $n$th eigenvalue is asymptotic to $A\sqrt{n}$ as $n\to\infty$, where
$A=2^{5/6}$. In Sec.~\ref{s3} we perform an asymptotic calculation of $Z(t)$
correct to order $1/\lambda$ and use this result to obtain the number $A$ in
(\ref{e9}). In Sec.~\ref{s4} we suggest that the techniques presented in this
paper may apply to many other nonlinear differential equations. As evidence, we
present numerical results regarding the first Painlev\'e transcendent. We also
conjecture that the number $A$ in (\ref{e9}) may be related to the power-series
constant $P$, which describes the asymptotic behavior of the zeros of partial
sums of Taylor series of analytic functions.

\section{Numerical study of (\ref{e1}) and (\ref{e14})}
\label{s2}

We begin our analysis of (\ref{e1}) by constructing the Taylor series expansion 
\begin{equation}
y(x)=\sum_{n=0}^\infty b_nx^n
\label{e15}
\end{equation}
of the solution $y(x)$. To find the Taylor coefficients $b_n$ we substitute this
expansion into the differential equation and collect powers of $x$. The first
few Taylor coefficients are
\begin{eqnarray}
b_0&=&y(0)=a,\nonumber\\
b_1&=&1,\nonumber\\
b_2&=&0,\nonumber\\
b_3&=&-\textstyle{\frac{1}{6}}\pi^2a^2,\nonumber\\
b_4&=&-\textstyle{\frac{1}{4}}\pi^2a,\nonumber\\
b_5&=&\textstyle{\frac{1}{120}}\pi^4a^4-\textstyle{\frac{1}{10}}\pi^2,
\nonumber\\
b_6&=&\textstyle{\frac{1}{18}}\pi^4a^3,\nonumber\\
b_7&=&-\textstyle{\frac{1}{5040}}\pi^6a^6+\textstyle{\frac{2}{21}\pi^4a^2},
\nonumber\\
b_8&=&-\textstyle{\frac{1}{180}}\pi^6a^5+\textstyle{\frac{31}{480}}\pi^4a,
\nonumber\\
b_9&=&\textstyle{\frac{1}{362880}}\pi^8a^8-\textstyle{\frac{161}{6480}}\pi^6a^4
+\textstyle{\frac{17}{1080}}\pi^4.
\label{e16}
\end{eqnarray}

We then observe that we can reorganize and regroup the terms in the Taylor
series. For example, the first terms in $b_1$, $b_3$, $b_5$, $b_7$, $b_9$, and
so on, give rise to the function
$$\frac{1}{\pi a}\sin s$$
and the first terms in $b_4$, $b_6$, $b_8$, $b_{10}$, and so on, give rise to
$$\frac{1}{8\pi^2a^3}\left[2s\sin(2s)+\cos(2s)-2s^2-1\right],$$
where $s=\pi ax$. This partial summation of the Taylor series, a procedure used
in multiple-scale perturbation theory to eliminate secular behavior \cite{R7},
shows that the solution $y(x)$ is approximately a falling parabola with an
oscillatory contribution whose amplitude that is of order $1/a$. This is indeed
what we observe in Fig.~\ref{F1}. The partial summation of the Taylor series
suggests that $a$ and $y$ are both of order $\sqrt{n}$ and motivates the changes
of variable in (\ref{e12}) and (\ref{e13}), which give the scaled differential
equation (\ref{e14}).

\begin{figure}[h!]
\begin{center}
\includegraphics[trim=32mm 85mm 32mm 85mm,clip=true,scale=0.90]{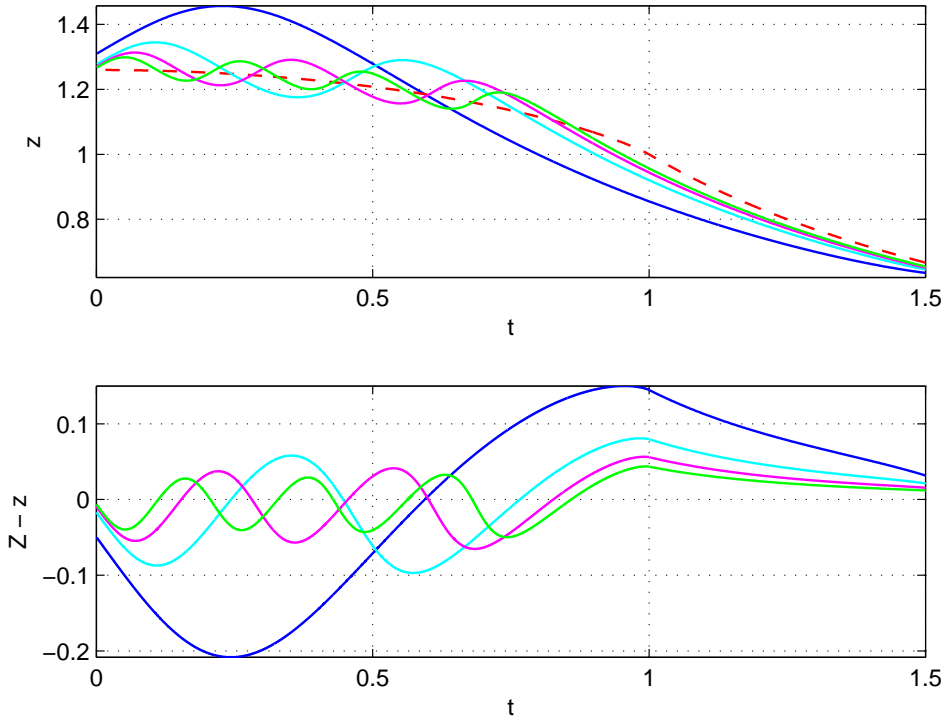}
\end{center}
\caption{Upper panel: Numerical plots of the first four separatrix solutions
$z(t)$ (eigensolutions) to (\ref{e14}) (blue, cyan, magenta, and green in the
electronic version). These solutions have one, two, three, and four maxima. As
$\lambda$ increases, these curves approach the solution to (\ref{e14}) for
$\lambda=\infty$ (dashed curve) (red in the electronic version). [The $\lambda=
\infty$ curve is called $Z(t)$ and satisfies the differential equation 
(\ref{e31}).] Lower panel: A plot of the differences between the solid curves
and the dashed curve.}
\label{F3}
\end{figure}

As $\lambda$ in (\ref{e14}) tends to $\infty$, the oscillations disappear.
(This is demonstrated in Sec.~\ref{s3}.) The resulting curve $Z(t)$, which
begins at $Z(0)=2^{1/3}$ and passes through $Z(1)=1$, is shown as a dashed line
(red in the electronic version) in Fig.~\ref{F3} (upper panel). Also shown are
the first four eigencurve (separatrix) solutions to (\ref{e14}) (blue, cyan,
magenta, and green in the electronic version), which have one, two, three, and
four maxima. Note that these eigensolutions rapidly approach the limiting dashed
curve as the number of oscillations increases. The lower panel in Fig.~\ref{F3}
indicates the difference between the dashed curve and the solid curves plotted
in the upper panel.

\begin{figure}[h!]
\begin{center}
\includegraphics[trim=32mm 85mm 32mm 85mm,clip=true,scale=0.90]{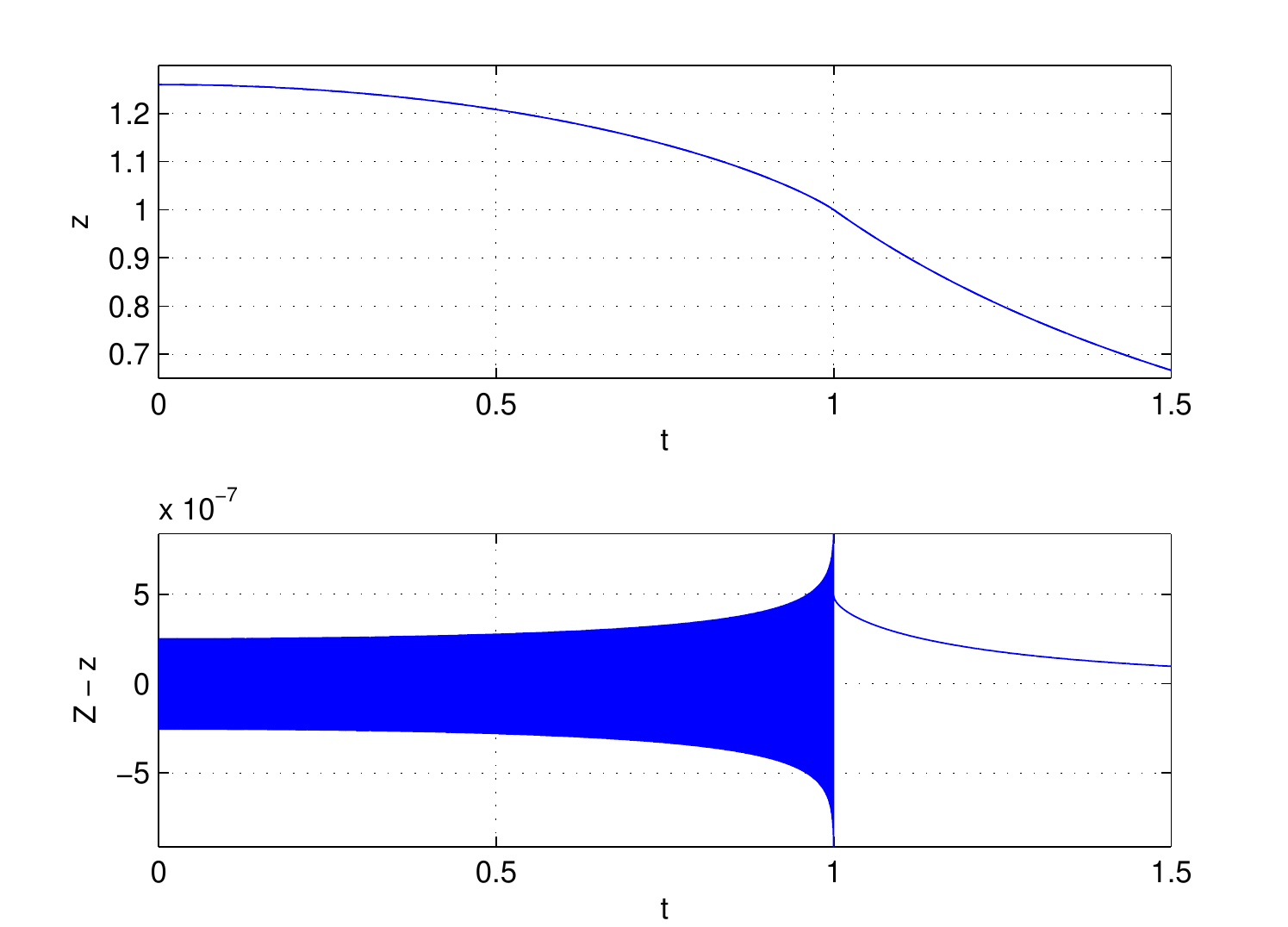}
\end{center}
\caption{Upper panel: Numerical solution $z(t)$ to (\ref{e14}) corresponding to
$n=500,000$. No oscillation is visible because the amplitude of oscillation is
of order $1/\lambda$ when $\lambda$ is large. Lower panel: Difference between
the $n=500,000$ eigencurve $z(t)$ and the $\lambda=\infty$ curve $Z(t)$. Note
that the difference is highly oscillatory and is of order $10^{-6}$.}
\label{F4}
\end{figure}

For large values of $\lambda$ the convergence to the limiting curve is dramatic.
In Fig.~\ref{F4} we plot the limiting curve $Z(t)$ in the upper panel and the
difference between the limiting curve and the $n=500,000$ separatrix curve
(eigencurve) in the lower panel. Note that the difference is of order $1/n$ 
($10^{-6}$). On the basis of these numerical calculations we were able to use
Richardson extrapolation \cite{R8} to calculate the coefficient $A$ accurate to
one part in $10^{10}$ and we conjectured reliably that $A=2^{5/6}$.

\begin{figure}[h!]
\begin{center}
\includegraphics[trim=0mm 70mm 0mm 70mm,clip=true,scale=.70]{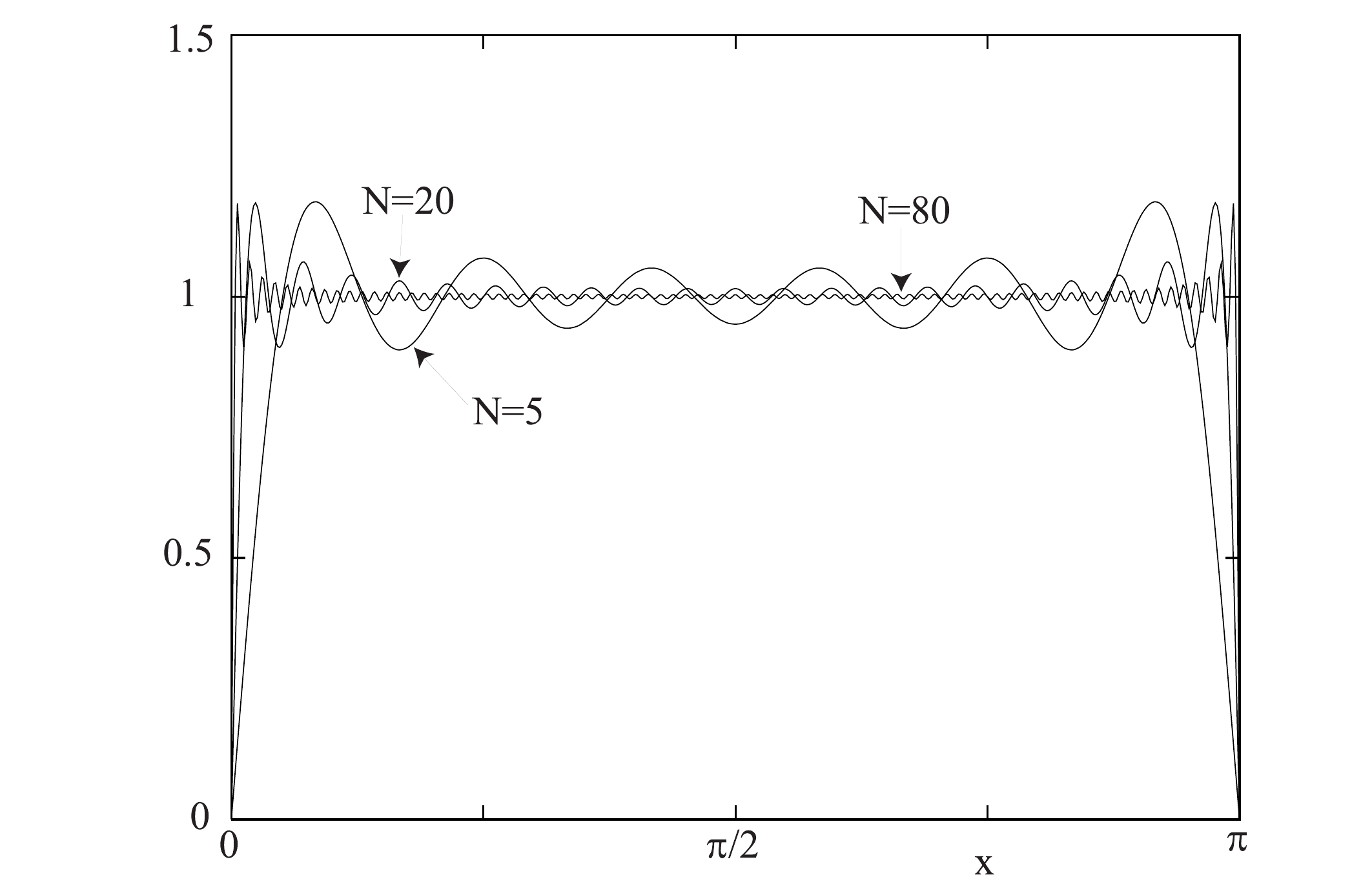}
\end{center}
\caption{Convergence of the $N=5$, $20$, and $80$ partial sums in (\ref{e17}) of
the Fourier sine series for $f(x)=1$. The partial sums of the Fourier series
converge to $1$ as $N\to\infty$ in much the same way that $z(t)$ converges to
$Z(t)$ as $\lambda\to\infty$. As $N$ increases, the frequency of oscillation
increases and the amplitude of oscillation approaches zero.}
\label{F5}
\end{figure}

The convergence of $z(t)$ (which is rapidly oscillatory when $0\leq t\leq1$) to
$Z(t)$ (which is smooth and nonoscillatory) as $\lambda\to\infty$ strongly
resembles the convergence of a Fourier series. Consider, for example, the
convergence of the Fourier sine series to the function $f(x)=1$ on the interval
$0<x<\pi$. The $2N+1$ partial sum of the Fourier sine series is
\begin{equation}
S_{2N+1}(x)=\frac{4}{\pi}\sum_{n=0}^N\frac{\sin[(2n+1)x]}{2n+1}.
\label{e17}
\end{equation}
As can be inferred from Fig.~\ref{F5}, which displays the partial sums for $N=5,
\,20,\,80$, as $N$ increases, $S_{2N+1}(x)$ approaches 1 (except for values of
$x$ near $x=0$ and $x=\pi$) in a highly oscillatory fashion that strongly
resembles the approach of $z(t)$ to $Z(t)$ in Fig.~\ref{F4}.

\section{Asymptotic solution of the scaled equation (\ref{e14})}
\label{s3}

The objective of the asymptotic analysis in this section is to solve (\ref{e14})
for large $\lambda$ and to verify the result in (\ref{e9}); namely, that $A=2^{
5/6}$. We begin by converting the differential equation in (\ref{e14}) to the
integral
equation
\begin{equation}
[z(t)]^2-[z(0)]^2+t^2/2+\eta(t)={\rm O}(1/\lambda)\quad(\lambda\to\infty),
\label{e18}
\end{equation}
where
\begin{equation}
\eta(t)=\int_0^t ds\,s\cos[2\lambda sz(s)].
\label{e19}
\end{equation}
To obtain this result we multiply (\ref{e14}) by $z(t)+tz'(t)$, integrate from
$0$ to $t$, and use the double-angle formula for the cosine function.

The problem is now to calculate $\eta(t)$. To do so, we observe that $\eta(t)$
is just one of an infinite set of moments $A_{n,k}(t)$, which are defined as
follows:
\begin{equation}
A_{n,k}(t)\equiv\int_0^t ds\cos[n\lambda sz(s)]\frac{s^{k+1}}{[z(s)]^k}.
\label{e20}
\end{equation}
Note that $\eta(t)=A_{2,0}(t)$.

For large $\lambda$ these moments satisfy the linear difference equation
\begin{equation}
A_{n,k}(t)=-\half A_{n-1,k+1}(t)-\half A_{n+1,k+1}(t)\quad(n\geq2).
\label{e21}
\end{equation}
To obtain this equation we multiply the integrand of the integral in (\ref{e20})
by
\begin{equation}
\frac{z(s)+sz'(s)}{z(s)}-\frac{sz'(s)}{z(s)}.
\label{e22}
\end{equation}
(Note that this quantity is merely an elaborate way of writing $1$.) We then
evaluate the first part of the resulting integral by parts and verify that it is
negligible as $\lambda\to\infty$ if $t\leq1$. In the second part of the integral
we replace $z'(t)$ by $\cos[\lambda tz(t)]$ and use the trigonometric identity
$$\cos(na)\cos(a)=\half\cos[(n+1)a]+\half\cos[(n-1)a].$$

By using repeated integration by parts it is easy to show that $\eta(t)$ in
(\ref{e19}) can be expanded as the series
\begin{equation}
\eta(t)=\sum_{p=0}^\infty \alpha_{1,2p+1}A_{1,2p+1}(t),
\label{e23}
\end{equation}
where the coefficients $\alpha_{n,k}$ are determined by a one-dimensional
random-walk process in which random walkers move left or right with equal
probability but become static when they reach $n=1$. The initial condition for
the random walk is that $\alpha_{n,0}=0$ if $n\neq2$ and $\alpha_{2,0}=1$. The
coefficients $\alpha_{n,k}$ obey the difference equations
\begin{equation}
2\alpha_{1,k}+\alpha_{2,k-1}=0,
\label{e24}
\end{equation}
\begin{equation}
2\alpha_{2,k}+\alpha_{3,k-1}=0,
\label{e25}
\end{equation}
\begin{equation}
2\alpha_{n,k}+\alpha_{n-1,k-1}+\alpha_{n+1,k-1}=0\quad(n\geq3).
\label{e26}
\end{equation}
(Note that $\alpha_{n,k}=0$ if one of the subscripts is odd and the other is
even.) The difference equations (\ref{e25}) and (\ref{e26}) can be solved in
closed form, and we obtain the following exact result for $n\geq2$:
\begin{equation}
\alpha_{n,k}=\frac{(-1)^n(n-1)k!}{2^k(k/2+n/2)!(k/2-n/2+1)!},
\label{e27}
\end{equation}
which holds if $n$ and $k$ are both even or both odd. Finally, we use equation
(\ref{e24}) to obtain 
\begin{equation}
\alpha_{1,2p+1}=-\half\alpha_{2,2p}=-\frac{(2p)!}{2^{2p+1}p!(p+1)!}=-\frac{
\Gamma(p+1/2)}{2\sqrt{\pi}\,(p+1)!},
\label{e28}
\end{equation}
where the duplication formula for the Gamma function was used to obtain the
last equality.

Thus, the series in (\ref{e23}) for $\eta(t)$ reduces to the series of integrals
$$\eta(t)=-\frac{1}{2\sqrt{\pi}}\sum_{p=0}^\infty\frac{\Gamma(p+1/2)}{(p+1)!}
\int_0^t ds\,z'(s)\frac{s^{2p+2}}{[z(s)]^{2p+1}},$$
which is valid for $t\leq1$. This series can be summed in closed form:
\begin{equation}
\eta(t)=\int_0^t ds\,z(s)z'(s)\sqrt{1-s^2/[z(s)]^2}-\int_0^t ds\,z(s)z'(s).
\label{e29}
\end{equation}
There is no explicit reference to $\lambda$ in this expression, so we pass to
the limit as $\lambda\to\infty$. In this limit the function $z(t)$, which is
rapidly oscillatory (see Fig.~\ref{F4}), approaches the function $Z(t)$, which
is smooth and not oscillatory. We therefore obtain from (\ref{e18}) an integral
equation satisfied $Z(t)$:
\begin{equation}
[Z(t)]^2-[Z(0)]^2+\half t^2-\int_0^t ds\,Z(s)Z'(s)+\int_0^t ds\,Z(s)Z'(s)\sqrt{
1-s^2/[Z(s)]^2}=0.
\label{e30}
\end{equation}
We differentiate (\ref{e30}) to obtain an elementary differential equation
satisfied by $Z(t)$:
\begin{equation}
Z(t)Z'(t)+t+Z'(t)\sqrt{[Z(t)]^2-t^2}=0.
\label{e31}
\end{equation}

This differential equation is easy to solve because it is of {\it homogeneous}
type; that is, the equation can be rearranged so that $Z(t)$ is always
accompanied by a factor of $1/t$. Such an equation can be solved by making the
substitution $Z(t)=tG(t)$ to reduce (\ref{e31}) to a separable differential
equation for $G(t)$. The general solution for $G(t)$ is
\begin{equation}
\frac{K}{t^3}=\left(1+3[G(t)]^2\right)\left(G(t)+\sqrt{[G(t)]^2-1}\right)
\frac{\sqrt{[G(t)]^2-1}-2G(t)}{\sqrt{[G(t)]^2-1}+2G(t)},
\label{e32}
\end{equation}
where $K$ is an arbitrary constant. The condition that $G(1)=1$ then determines 
that $K=-4$, and we obtain the exact result that $Z(0)=2^{1/3}$. We thus
conclude that $A=2^{5/6}$. This establishes the principal result of this paper.

\section{Discussion and description of future work}
\label{s4}

\subsection{First Painlev\'e transcendent}
\label{ss4A}
We believe that the asymptotic approach developed in this paper may be
applicable to many nonlinear differential equations having separatrix structure.
One possibility is the differential equation for the first Painlev\'e
transcendent
\begin{equation}
y''(x)=[y(x)]^2+x.
\label{e33}
\end{equation}
How do solutions to this equation behave as $x\to-\infty$? It is clear that
when $x$ becomes large and negative, there can be a dominant asymptotic balance
between the positive term $[y(x)]^2$ and the negative term $x$, which implies
that $y(x)$ can have two possible leading asymptotic behaviors:
\begin{equation}
y(x)\sim\pm\sqrt{-x}\quad(x\to-\infty).
\label{e34}
\end{equation}
This asymptotic result is justified because the second derivative of $\sqrt{-x}$
is small compared with $x$ when $|x|$ is large.

This problem is interesting because the asymptotic behavior $y(x)\sim-\sqrt{-x}$
is stable but the asymptotic behavior $y(x)\sim\sqrt{-x}$ is unstable. To verify
this, we calculate the corrections to these two asymptotic behaviors. It is easy
to show that when $x$ is large and negative, the solution to (\ref{e33})
oscillates about and decays slowly towards the curve $-\sqrt{-x}$ \cite{R1}:
\begin{equation}
y(x)\sim-\sqrt{-x}+c(-x)^{-1/8}\cos\left[\textstyle{\frac{4}{5}}\sqrt{2}(-x)^{
5/4}+d\right]\quad(x\to-\infty),
\label{e35}
\end{equation}
where $c$ and $d$ are two arbitrary constants. The differential equation
(\ref{e33}) is second order and, as expected, this asymptotic behavior contains
two arbitrary constants.

On the other hand, the correction to the $+\sqrt{-x}$ behavior has an 
exponential form
\begin{equation}
y(x)\sim\sqrt{-x}+c_\pm(-x)^{-1/8}\exp\left[\textstyle{\pm\frac{4}{5}}\sqrt{2}(
-x)^{5/4}\right]\quad(x\to-\infty).
\label{e36}
\end{equation}
Thus, if $c_+\neq0$, nearby solutions generally veer away from the curve
$\sqrt{-x}$ as $x\to-\infty$. The special solutions that decay exponentially
towards the curve $\sqrt{-x}$ form a one-parameter and not a two-parameter class
because $c_+=0$. The vanishing of the parameter $c_+$ gives rise to an
eigenvalue condition on the choice of initial value of $y'(0)$. For each value
of $y(0)$ there is a set of eigencurves (separatrices). These curves
correspond to a discrete set of initial slopes $y'(0)$.

\begin{figure}[h!]
\begin{center}
\includegraphics[trim=5mm 55mm 5mm 55mm,clip=true,scale=.70]{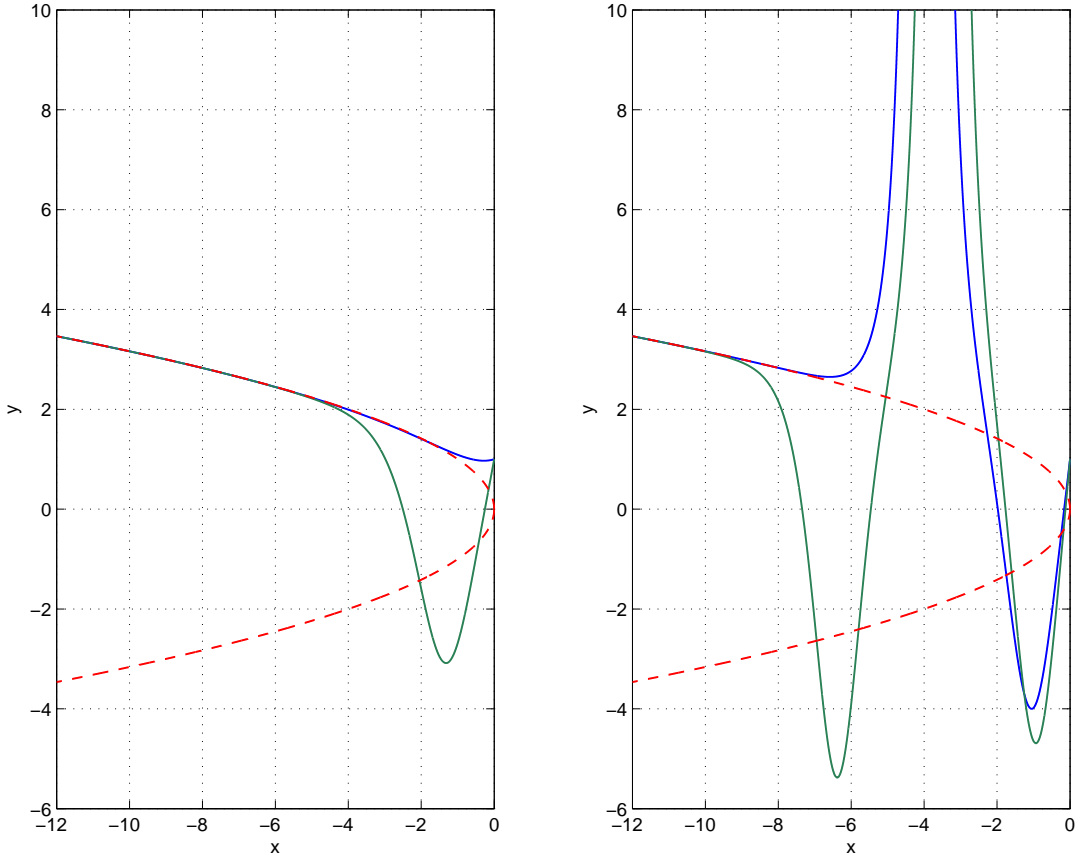}
\end{center}
\caption{Eigencurve solutions to the first Painlev\'e trenscendent. The
eigencurves pass through $y(0)=1$ and the slopes of the curves at $x=0$ are the
eigenvalues $a_n$. As $x\to-\infty$, the eigencurves approach $+\sqrt{-x}$
exponentially rapidly. Left panel: first two eigencurves corresponding to the
eigenvalues $a_1=0.231955$ and $a_2=3.980669$. The $a_1$ curve approaches $+
\sqrt{-x}$ from above and the $a_2$ curve approaches $+\sqrt{-x}$ from below.
Right panel: The second two eigencurves for the Painlev\'e transcendent
corresponding to the eigenvalues $a_3=6.257998$ and $a_4=8.075911$. Note that
the second pair of eigenvalues passes through one double pole before approaching
the curve $+\sqrt{-x}$.}
\label{F6}
\end{figure}

\begin{figure}[h!]
\begin{center}
\includegraphics[trim=5mm 55mm 5mm 55mm,clip=true,scale=.70]{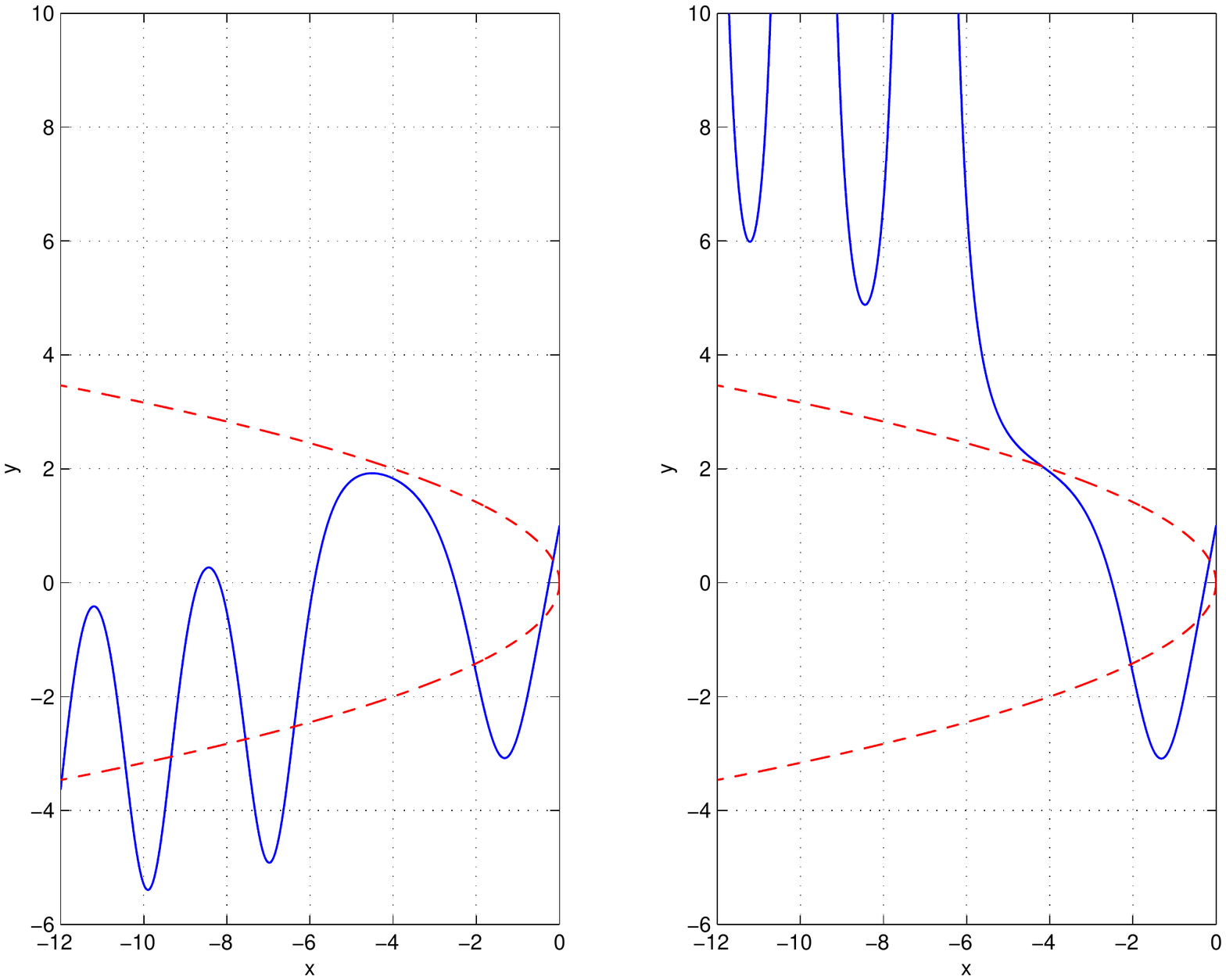}
\end{center}
\caption{Non-eigenvalue solutions to the first Painlev\'e transcendent. If
$y(0)=1$ but $y'(0)$ is not one of the eigenvalues $a_n$, the curve either
oscillates about and approaches the stable asymptotic curve $-\sqrt{-x}$ as in
the left panel or else it lies above the unstable asymptotic curve $+\sqrt{-x}$
and passes through an infinite number of double poles as in the right panel.}
\label{F7}
\end{figure}

We have performed a numerical study of the solutions to (\ref{e33}) that satisfy
the initial conditions $y(0)=1$ and $y'(0)=a$. There is a discrete set of
eigencurves whose initial positive slopes are $a_1=0.231955$, $a_2=3.980669$,
$a_3=6.257998$, $a_4=8.075911$, $a_5=9.654843$, $a_6=11.078201$, $a_7=
12.389217$, $a_8=13.613878$, $a_9=14.769304$, $a_{10}=15.867511$, $a_{11}=
16.917331$, $a_{12}=17.925488$. (There is also an infinite discrete set of {\it
negative} eigenvalues.) The first two of these curves are shown in the left
panel and the next two of these curves are shown in the right panel of
Fig.~\ref{F6}. Note that the separatrix curves do not just exhibit $n$ maxima
as do the dashed curves in Fig.~\ref{F2}. Rather, these curves pass through
increasingly many double poles. The curve corresponding to $a_1$ approaches
$+\sqrt{-x}$ from above and the curve corresponding to $a_2$ approaches $+\sqrt{
-x}$ from below. The curves corresponding to $a_3$ and $a_4$ also approach $+
\sqrt{-x}$ from above and below, but these curves both pass through one double
pole. Similarly, the curves corresponding to $a_5$ and $a_6$ pass through two
double poles, and the curves corresponding to $a_{2n-1}$ and $a_{2n}$ pass
through $n$ double poles. The key feature of these separatrix curves is that
after passing through $n$ double poles, they approach the curve $+\sqrt{-x}$
exponentially fast as $x\to-\infty$. If the value of $y'(0)$ lies in between two
eigenvalues, the curve either oscillates about and approaches the stable
asymptotic curve $-\sqrt{-x}$ as in the left panel of Fig.~\ref{F7} or else it
lies above the unstable asymptotic curve $+\sqrt{-x}$ and passes through an
infinite number of double poles as in the right panel of Fig.~\ref{F7}.

We have used Richardson extrapolation \cite{R8} to find the behavior of the
numbers $a_n$ for large $n$, and we obtain a result very similar in structure
to that in (\ref{e9}). Specifically, we find that 
\begin{equation}
a_n\sim Cn^{3/5}\quad(n\to\infty),
\label{e37}
\end{equation}
where $C=4.28373$. This number is very close to $\frac{17}{5}2^{1/3}$. The
constant $C$ appears to be universal in that it is seems to be the same for all
values of $y(0)$. We are currently trying to apply our analytical asymptotic
methods to this problem to find an analytic calculation for the number $C$.

\subsection{Conjectural connection with the power-series constant}
\label{ss4B}
In conclusion, we point out a possible connection between this work and the
power-series constant $P$ in the theory of complex variables. (The quest to
find the value of the power series constant was originated by Hayman \cite{R9}.)
Let $\cF$ be the class of functions $f(z)$ that are analytic for $|z|<1$ but not
analytic for $|z|<r$ for $r>1$. If $f\in\cF$, its power-series expansion
$$f(z)=\sum_{k=0}^\infty a_k z^k$$
has a radius of convergence of $1$. We denote $\rho_n(f)$ as the largest number
$r$ such that the partial sum of $f(z)$,
\begin{equation}
S_n(z)=\sum_{k=0}^n a_kz^k,
\label{e38}
\end{equation}
has at least one zero on $|z|=r$. We define
\begin{equation}
\rho(f)\equiv\lim_{n\to\infty}\inf\rho_n(f).
\label{e39}
\end{equation}
The power series constant $P$ is then defined as
\begin{equation}
P\equiv\sup_{f\in\cF}\rho(f).
\label{e40}
\end{equation}

The precise value for $P$ is not known. However, lower and upper bounds on $P$
have been established. The power series constant was known to lie in the
interval $1\leq P\leq 2$ until Clunie and Erd\"os \cite{R10} sharpened these
bounds to $\sqrt{2}\leq P\leq 2$. Buckholtz \cite{R11} further sharpened these
bounds to $1.7\leq P\leq 12^{1/4}$, which was optimized by Frank \cite{R11} to
\begin{equation}
1.7818\leq P\leq 1.82.
\label{e41}
\end{equation}
The latter values appear to be the best known values to date. It is astonishing
that the value of $A$ in (\ref{e9}) agrees with the best known lower bound for
the value of $P$. We do not know whether our value $2^{5/6}$ coincides exactly
with the lower bound. We leave this observation here as coincidence and hope to
elaborate on the precise relation in a future paper \cite{R12}.

\begin{figure}[h!]
\begin{center}
\includegraphics[scale=.35]{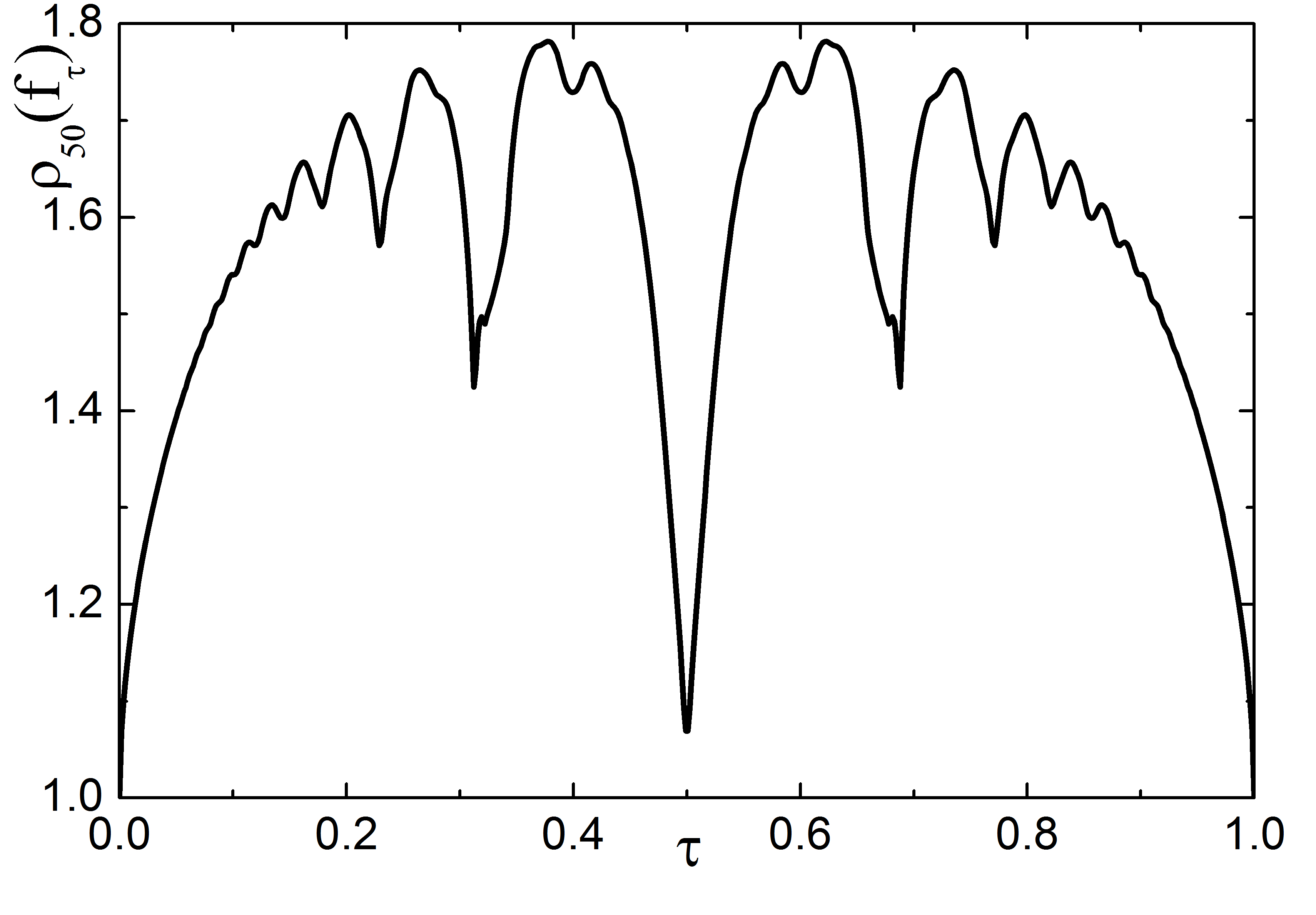}
\end{center}
\caption{A plot of $\rho_{50}(f_\tau)$ as a function of $\tau$. Note that at
the optimal value of the parameter $\tau$, the maximum of the curve is very
close to the value $1.7818$.}
\label{F8}
\end{figure}

As an illustration, let us compute $\rho(f)$ for some specific functions. In
some rare cases we can sum the entire series and can therefore compute the value
exactly. For instance, for the class of functions
\begin{equation}
f_\tau(z)=\sum_{k=0}^\infty\exp[i\pi\tau(k^2+k)]~z^k
\label{e42}
\end{equation}
we compute
\begin{equation}
f_{1/4}(z)=\frac{1+iz-iz^2-z^3}{1+z^4}
\label{e43}
\end{equation}
with zeros $z_1=1$ and $z_{2/3}=-1/2(1+i)\pm\sqrt{2i-4}$, such that $\rho(f_{1/4
})=\left\vert z_2\right\vert\approx1.70002$. We can also compute a value closer
to $P$:
\begin{equation}
f_{3/8}(z)=\frac{1+e^{\frac{3i\pi}{4}}z+e^{\frac{i\pi}{4}}z^2+iz^3-iz^4
-e^{\frac{i\pi }{4}}z^5-e^{\frac{3i\pi}{4}}z^6-z^7}{1+z^8},
\label{e44}
\end{equation}
leading to $\rho\left(f_{3/8}\right)\approx1.7804$. We must terminate the
summation and evaluate $\rho_n(f)$ for a sufficiently large value of $n$. In
Fig.~\ref{F8} we display our numerical results for $\rho_{50}(f_\tau)$ obtained
from the partial sum $S_{50}(z)$. The maximum values are $\rho_{50}(f_{0.3780})=
\rho_{50}(f_{0.8780})\approx1.7818$, which coincide with the best known lower
bound for $P$ up to the precision of the computation.

\subsection{Final comments}
\label{ss4C}

In this paper we have focused on separatrix behavior, which is a consequence of
instabilities of nonlinear differential equations. We have interpreted 
separatrices as being eigenfunctions (eigencurves). The corresponding
eigenvalues are the initial conditions needed to specify the separatrix curves.
For the differential equation $y'(x)=\cos[\pi xy(x)]$, we have shown that the
$n$th eigenvalue grows like $2^{1/3}\sqrt{2n}$ for large $n$. The number
$2^{1/3}$ appears again in the numerical analysis of the first Painlev\'e
transcendent. Moreover, to the currently known precision, the number $2^{5/6}$
appears in another asymptotic context, namely, as the lower bound $1.7818$ on
the power series constant $P$. We conjecture that the number $2^{5/6}$ may even
be the exact value of $P$.

We emphasize that in this paper we have been interested in the limit of large
eigenvalues. For linear eigenvalue problems this limit is accessible by using
WKB theory. In the case of the nonlinear eigenvalue problem solved in this paper
the large-eigenvalue limit is accessible because the problem becomes {\it
linear} in this limit; indeed, the large-eigenvalue separatrix curve was found
by reducing the problem to a {\it linear} random walk problem, which can be
solved exactly. The strategy of transforming a nonlinear problem to an
equivalent linear problem is reminiscent of the Hopf-Cole substitution that
reduces the nonlinear Burgers equation to the linear diffusion equation or the
inverse-scattering analysis that reduces the nonlinear Korteweg-de Vries
equation to a linear integral equation.

There is a plausible argument that the power series constant $P$ is connected
with the asymptotic behavior of eigenvalues: On one hand, $P$ is associated with
the zero of largest modulus of a polynomial of degree $n$, which is constructed
as the $n$th partial sum of a Taylor series. On the other hand, a conventional
linear eigenvalue problem of the form $H\psi=E\psi$ may be solved by introducing
a basis and replacing the operator $H$ by an $N\times N$ matrix $H_N$. Then, to
calculate the eigenvalues numerically we find the zeros of the secular
polynomial ${\rm Det}(H_N-IE)$. Finding the asymptotic behavior of the
high-energy eigenvalues corresponds to finding the behavior of the largest zero
of the secular polynomial as $N$, the degree of the polynomial, tends to
infinity.

We believe that the techniques proposed here to find the asymptotic behavior of
large eigenvalues may extend to other nonlinear differential equations
exhibiting instabilities and separatrix behavior.

\acknowledgments
CMB and JK thank the U.S. Department of Energy for financial support.

\end{document}